\documentclass[manuscript]{aastex}
\usepackage{graphicx}

\shorttitle{Diagnostic tools for X-ray observations}
\shortauthors{Garc\'{\i}a \and Kallman}

\begin{document}

\title{X-ray reflected spectra from accretion disk models. \\
II. Diagnostic tools for X-ray observations}

\author{J.~Garc\'ia\altaffilmark{1,2},
       T.R.~Kallman\altaffilmark{2},
       \and R.F.~Mushotzky\altaffilmark{3}} 
\altaffiltext{1}{Department of Physics, Western Michigan University, 
Kalamazoo, MI 49008, USA \email{javier.garcia@wmich.edu}}
\altaffiltext{2}{NASA Goddard Space Flight Center, Greenbelt, MD 20771 
\email{timothy.r.kallman@nasa.gov}}
\altaffiltext{3}{Department of Astronomy, University of Maryland, College Park, MD, USA
\email{richard@astro.umd.edu}}
%
%==================================================================================
%
\begin{abstract}
We present a comprehensive study of the emission spectra from accreting sources.
We use our new reflection code to compute the reflected spectra from an accretion
disk illuminated by X-rays. This set of models covers different values of ionization  
parameter, solar iron abundance and photon index for the illuminating spectrum. 
These models also include the most complete and recent atomic data for the inner-shell 
of the  iron and oxygen isonuclear sequences. We concentrate our analysis to the
$2-10$~keV energy region, and in particular to the iron K-shell emission lines.
We show the dependency of the equivalent width (EW) of the Fe K$\alpha$ with the 
ionization parameter. The maximum value of the EW is $\sim 800$~eV for models
with log~$\xi\sim 1.5$, and decreases monotonically as $\xi$ increases. For lower values
of $\xi$ the Fe K$\alpha$ EW decreases to a minimum near log~$\xi\sim 0.8$.
We produce simulated CCD observations based on our reflection models. For
low ionized, reflection dominated cases, the $2-10$~keV energy region shows a very
broad, curving continuum that cannot be represented by a simple power-law. 
We show that in addition to the Fe K-shell emission, there are other prominent 
features such as the Si and S L$\alpha$ lines, a blend of Ar~{\sc viii-xi} lines,
and the Ca~{\sc x} K$\alpha$ line. In some cases the S~{\sc xv} blends with the
He-like Si RRC producing a broad feature that cannot be reproduced by a simple
Gaussian profile. This could be used as a signature of reflection.
%A comparison of the line centroid energies versus the equivalent width of the 
%Fe K$\alpha$ predicted by our models and those derived from Seyfert galaxies
%and low mass X-ray binaries (LMXB) shows good agreement. Only models with low ionization
%(log~$\xi<2$) and with no contribution from the power-law (pure reflection), can
%achieve large values of the EW ($\sim 1$~keV). Only Seyfert 1.8-2 galaxies, or
%those with $N_{\mathrm{H}}<10^{21}$~cm$^{-2}$ show such values. 
%For the LMXB, EWs are around $\sim 10-200$~eV. Since all these sources show 
%hotter Fe K-lines (with energies around $6.7-6.9$~keV), they agree with models 
%with values of the ionization parameter larger than those for active galactic
%nuclei ($2 < \mathrm{log}~\xi < 4$).
\end{abstract}
%
%==================================================================================
%
\section{Introduction}
Accreting systems are observed to emit copious radiation in the X-ray energy
range which suggests emission from the innermost regions of an accretion
disk. Analysis of the X-ray spectra is crucial to study the complex
mixture of emitting and absorbing components in the circumnuclear
regions of these systems. However, there are only few observables that 
provide indication of the existence of an
accretion disk. These need to be understood in order to correctly
interpret the physics of these systems. In a general sense, the observed radiation
from accreting sources can be divided into: a thermal component, in the form of
a black body emitted at the surface of the disk with typical temperatures
of $\sim 0.01-10$~keV (for the mass range $\sim 10^8-10~M_{\sun}$); 
a coronal component in the form of a power law covering
energies up to $\sim 100$~keV, believed to arise from inverse Compton scattering
in a hot gas that lies above the disk; and a reflected component, resulting
from the interaction of some of the coronal X-ray photons and the optically
thick material of the disk. 
In the reflected component, the most prominent feature is the
iron K$\alpha$ emission line at $\sim 6.4$~keV, produced by transitions of electrons
between the $1s$ and $2p$ atomic orbitals. These are ubiquitous in the spectra
of accreting sources \citep{pou90, nan94,mil07}. Other reflection signatures are
the so called Compton shoulder (next to the Fe K-line), and the Compton hump
(above $\sim 10$ keV), produced by the down-scattering of high energy photons by cold electrons.

Much theoretical effort has gone into studies of X-ray illuminated disks over 
the past few decades. Most models assume that the gas density 
is constant with depth \citep{don92,ros93,mat93,cze94,kro94,mag95,ros96,mat96,pou96,bla99}.
Although constant density models may be appropriate for radiation-pressure
dominated disks, other studies have shown significant differences when the
gas density is properly solved via hydrostatic equilibrium \citep{roz96,nay00, nay01, bal01, dum02, ros07}.
Recently we have developed a new model for the reflected spectra from illuminated accretion
disks called {\sc xillver} \citep{gar10}. Although our code is similar in its principal assumptions to previous models, 
{\sc xillver} includes the most recent and complete atomic data
for K-shell of all relevant ions \citep{kal04,gar05,pal08a,gar09a,pal10}. This has a
dramatic impact on the predicted spectra, in particular the K$\alpha$ emission from iron.
With this model we can study the effects of incident X-rays on the
surface of the accretion disk by solving simultaneously the equations of radiative
transfer and ionization equilibrium over a large range of column
densities. Plane-parallel geometry and azimuthal symmetry are assumed, such that
each calculation corresponds to an annular ring at a given distance from the 
source of X-rays. The redistribution of photons due to Compton down-scattering
is included by using a Gaussian approximation for the Compton kernel. With {\sc xillver}
we are able to solve the reflection problem with great detail, i.e., with very high energy,
spatial and angular resolution.

In this paper we present a systematic analysis of our models for reflected spectra from
X-ray illuminated accretion disks. We show how the most relevant atomic features
in the spectra depend on the assumed properties of the irradiated gas. We pay 
particular attention to the Fe K-shell emission lines, and we quantify its strength
in terms of the equivalent widths predicted by our models. These models are also
used to produce faked CCD spectra in order to simulate the effects of the instrumental
response and limited spectral resolution. These results will be helpful diagnostic
tools in the interpretation of accreting sources observations. 

In the next Section we describe briefly the numerical methods used in our 
reflection code. In Section~\ref{secres} we present the results of our analysis on the
simulated spectra, as well as comparisons with observations from Seyfert galaxies
and X-ray binaries. The main conclusions are presented in Section~\ref{seccon}
%
%==================================================================================
%
\section{The Reflection Model}\label{secmod}
In order to calculate the reflected spectra from X-ray illuminated accretion
disks we make use of our reflection code {\sc xillver}. The details of the calculations
are fully described in \cite{gar10}, thus here we just review the main aspects.

The description of the interaction of the radiation with the gas in the illuminated
slab requires the solution of the transfer equation: 
\begin{equation}\label{eqrt}
\mu^2\omega^2(E,\tau)\frac{\partial^2 u(\mu,E,\tau)}{\partial\tau^2} +
\mu^2\omega(E,\tau)\frac{\partial\omega(E,\tau)}{\partial\tau}
\frac{\partial u(\mu,E,\tau)}{\partial\tau} = u(\mu,E,\tau) - S(E,\tau)
\end{equation}
where $u(\mu,E,\tau)$ is the average intensity of the radiation field for a given
cosine of the angle with respect to the normal $\mu$, energy $E$, and position in
the slab, specified by the Thomson optical depth $d\tau \equiv -\alpha_{\mathrm{T}} dz =
- \sigma_{T} n_e dz$, where $\sigma_{T}$ is the Thomson cross section ($=6.65\times 10^{-25}$ cm$^2$),
and $n_e$ is the electron number density. The quantity $\omega$ is defined as
the ratio of the Thomson scattering coefficient $\alpha_{\mathrm{T}}$ to the total 
opacity $\chi(\tau,E)$. The second term in the right-hand side of Equation~\ref{eqrt}
is the source function, which is defined as the ratio of the total emissivity
over the total opacity of the gas, taking into account both scattering and
absorption processes:
\begin{equation}\label{eqsou}
S(E,\tau) = \frac{\alpha_{kn}(E)}{\chi(E,\tau)}J_c(E,\tau) + \frac{j(E,\tau)}{\chi(E,\tau)}
\end{equation}
where $\alpha_{kn}(E)$ is the Klein-Nishina scattering coefficient, $j(E,\tau)$ is
the thermal continuum plus lines emissivity, and $J_c(E,\tau)$ is the Comptonized mean
intensity, given by the Gaussian convolution
\begin{equation}\label{eqjc}
J_c(E,\tau) = \frac{1}{\sigma \pi^{1/2}} \int_0^{\infty}{dE' \exp{\left[ \frac{-(E-E_c)^2}{\sigma^2}\right]}
\int_0^1{u(\mu,E',\tau)d\mu}}
\end{equation}
The Gaussian is centered at $E_c = E'(1+4\theta-E'/m_ec^2)$, where $m_e$ is the electron mass,
$c$ is the speed of light, and $\theta=kT/m_ec^2$ is the dimensionless temperature.
The energy dispersion is given by $\sigma=E'\left[2\theta+\frac{2}{5}(E'/m_ec^2)^2\right]^{1/2}$.

The solution of the system is completed by imposing two boundary conditions. At the top
of the slab ($\tau=0$), we specify the incoming radiation field incident at a given angle $\mu_0$
by
\begin{equation}\label{eqbco}
\omega(0,E) \mu \left[ \frac{\partial u(\tau,\mu,E)}{\partial \tau} \right]_0
- u(0,\mu,E) = -\frac{2F_{\mathrm{x}}(E)}{\mu_0}\delta(\mu-\mu_0)
\end{equation}
where $F_{\mathrm{x}}(E)$ is the net flux of the illuminating radiation incident at the top of the slab. 
At the inner boundary ($\tau=\tau_{max}$), we specify the outgoing radiation field to be equal 
to a blackbody with the expected temperature for the disk:
\begin{equation}\label{eqbcm}
\omega(\tau_{max},\mu,E)\mu \left[ \frac{\partial u(\tau,\mu,E)}{\partial \tau} \right]_{\tau_{max}}
+ u(\tau_{max},\mu,E) = B(T_{disk})
\end{equation}
where $T_{disk}$ can be defined using the \cite{sak73} formulae. Since these models are
calculated under the assumption of constant density, we use the common definition of the
ionization parameter \citep{tar69} to characterize each case:
\begin{equation}\label{eqxi}
\xi = \frac{4\pi F_{\rm x}}{n_e}.
\end{equation}
where $n_e$ is the electron gas density, and $F_{\mathrm{x}}$ is the net illuminating flux 
integrated in the $1-1000$~Ry energy range. The solution of the system is found by forward 
elimination and back substitution. A full transfer solution must be achieved iteratively in order
to self-consistently treat the scattering process. This procedure requires $\sim \tau_{max}^2$
iterations (lambda iterations) for convergence.

Given the solution for the radiation field at each point in the atmosphere, 
we use the photoionization code {\sc xstar} \citep{kal01} to determine the 
state of the gas at each point of the gas. The state of the 
gas is defined by its temperature and the level populations of the ions. 
The relative abundances of the ions of a given element and the level populations 
are found by solving the ionization equilibrium equations under the assumption 
of local balance, subject to the constraint of particle number conservation for
each element.  {\sc xstar} calculates level populations, temperature, the 
opacity $\chi(E,\tau)$ and the total emissivity $j(E,\tau)$ of the gas assuming 
that all the physical processes are in steady state and imposing radiative 
equilibrium.

The {\sc xstar} atomic database collects recent data from many sources 
including CHIANTI \citep{lan06}, ADAS \citep{sum04}, NIST \citep{ral08}, 
TOPbase \citep{cun93} and the IRON project \citep{hum93}. The database
is described in detail by \cite{bau01}. Additionally, the atomic data 
associated with the K-shell of the Fe ions incorporated 
in the current version of {\sc xstar} has been recently calculated and
represents the most accurate and complete set available to the present.
A compilation of these results and a careful study of
their impact on the photoionization models can be found in \cite{kal04}.
Moreover, {\sc xstar} also includes the atomic data relevant to the 
photoabsorption near the K edge of all oxygen \citep{gar05}, and nitrogen
ions \citep{gar09a}.

Finally, all the calculations presented in this paper were carried out over 
a large optical depth ($\tau_{max}=10$), considering high resolution spectra 
with an energy grid of at least $5\times10^3$ points (${\cal R}=E/\Delta E\sim 350$),
200 spatial zones, and 50 angles to account for anisotropy of the radiation field.
The simulations do not take into account the dynamics of the system. 
Input parameters common to all these models are: the electron gas density $n_e=10^{15}$~cm$^{-3}$,
photon index of the incident radiation $\Gamma=2$, mass of the central object $M=10^8~M_{\sun}$, 
distance from the central object $R=7 R_s$, and the mass accretion rate
$\dot{M}=1.6\times 10^{-3} \dot{M}_{Edd}$, where $R_s=2GM/c^2$ is the Schwarzschild radius and
$\dot{M}_{Edd}$ is the accretion rate at the Eddington limit. This particular 
set of parameters yields a disk effective temperature of $T_{disk}=2.8\times 10^{4\ \circ}$K.
The models presented here cover 10 different values for the illumination 
flux $F_{\mathrm{x}}=5\times 10^{14}-5\times 10^{17}$~erg~cm$^{-2}$~s$^{-1}$, which
corresponds to ionization parameters of log~$\xi=0.8, 1.1, 1.5, 1.8, 2.1,...,3.8$.
It is important to notice that these simulations are not constrained to a
particular geometry for the illumination. This is because the flux $F_{\mathrm{x}}$
of the illuminating radiation is defined at the surface of the slab, regardless
of the position of the X-ray source. Therefore, a set of models with different
values of the ionization parameter can be used to construct a particular geometry,
by defining the luminosity and location of the source. We do not consider models with
ionization parameters lower than $\sim 4\pi$ (log~$\xi=0.8$), since our code is optimized
for calculations of medium to high ionization.
The upper limit is set to log~$\xi=3.8$, because at such a high illumination the
gas is almost completely ionized over a large depth.
%
%==================================================================================
%
\section{Results}\label{secres}
\subsection{Equivalent Widths}\label{secews}
The equivalent width of a line provides a quantitative measure of the strength of
the spectral profile, both in emission or absorption. It is defined by the
well known formula
\begin{equation}\label{eqew}
EW = \int_{E_{low}}^{E_{high}}{\frac{(F(E) - F_c(E))}{F_c(E)} dE},
\end{equation}
where $F(E)$ is the total flux and $F_c(E)$ is the total flux of the continuum under the line.
The integration is performed over the energy range where the spectral feature takes place,
between $E_{low}$ and $E_{high}$. Usually, there are uncertainties in
the determination of the intrinsic continuum which affects the knowledge of the integration
region and the actual value of the equivalent width itself. Nonetheless, one can approximate
its calculation by defining an energy region in which one knows only the spectral
feature of interest appears, and where a local continuum can be defined.

\begin{figure*}
\epsscale{1.0}\plotone{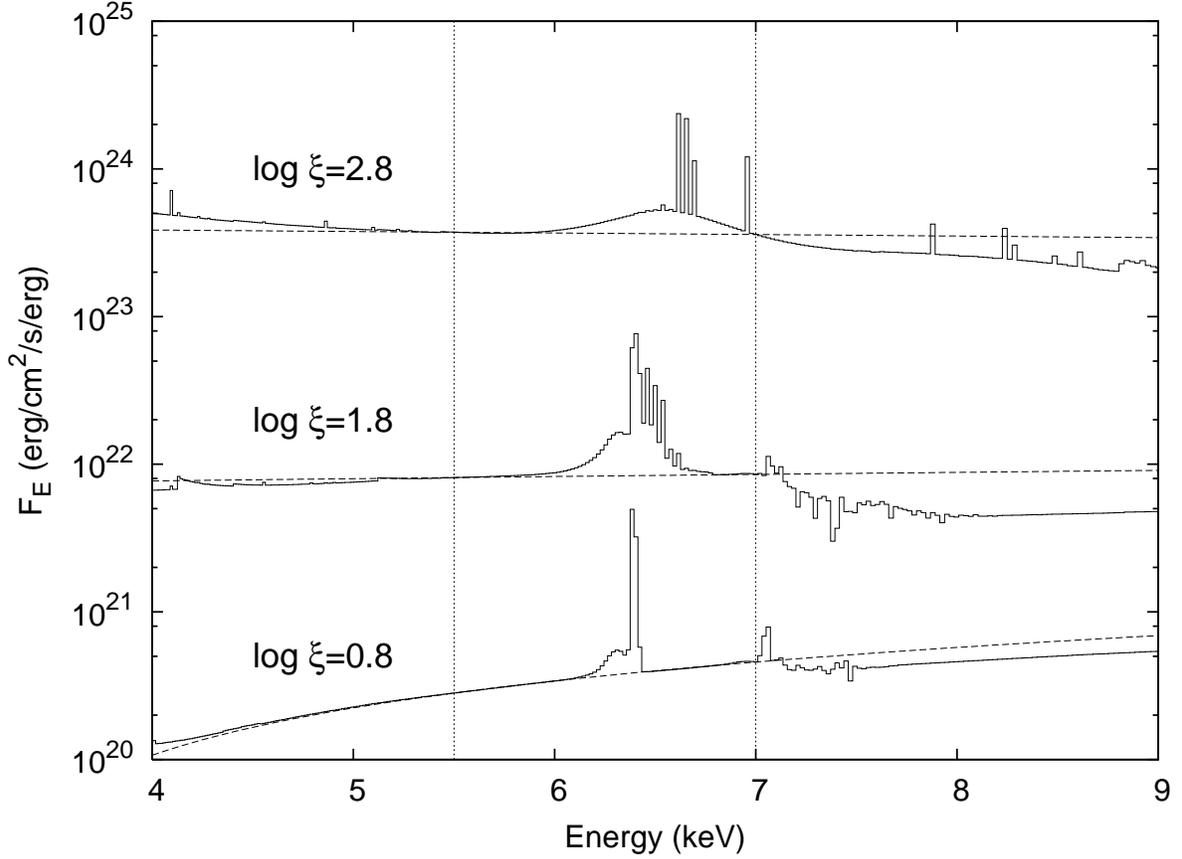}
\caption{Solid curves: emergent spectra from the surface of an accretion disk with 
constant density ($n=10^{15}$~cm$^{-3}$) calculated with our reflection code {\sc xillver}
for three different illuminations. The log of the ionization parameter is shown next
to each curve. The curves are rescaled for clarity. Dashed lines: local continuum used in
the calculation of the equivalent widths. Dotted vertical lines are placed at 5.5 and 7~keV.}
\label{f1}
\end{figure*}

In this paper we make use of the equivalent width as a measure of the strength of several
features in the X-ray spectra reflected from illuminated accretion disk. As an example, in
Figure~\ref{f1} we show the reflected spectra from 3 different models calculated for
log~$\xi=0.8, 1.8$ and $2.8$ in the 4-9 keV energy region, where the only atomic features
are due to inner-shell transitions from Fe ions. The spectra (flux vs. energy) are shown as
solid lines. Vertical dotted lines are placed at 5.5 keV and 7 keV, defining a particular
integration region. The continuum is defined as a straight line that passes through these two
points in the spectra, which is shown as dashed lines. It is clear from the Figure that
the resulting continuum does a good job reproducing the local continuum and that it is
not superimposed over any part of the emission profile. The upper limit $E_{high}=7$~keV
was chosen such that only emission from K$\alpha$ transitions of Fe are taken into account,
neglecting any K$\beta$ emission (which occurs at energies above 7~keV). The lower limit,
however, is more arbitrary. It needs to be chosen such that all the line profile is included
in the integration. If $E_{low}$ is set too large, part of the line emission can fall outside the
range, especially for cases with large values of the ionization parameter, where 
Comptonization smears the line profile due to the down-scattering. However,
if $E_{low}$ is set to a small value, all the line emission is taken into account but
the local continuum is modified to a point that could either under- or over-estimate 
the strength of the line. Therefore, we have repeated the calculation of the equivalent
widths for $E_{low}=5, 5.5$ and $6$~keV, in order to evaluate the sensitivity of the 
results on this lower limit. 

\begin{figure*}
\epsscale{1.0}\plotone{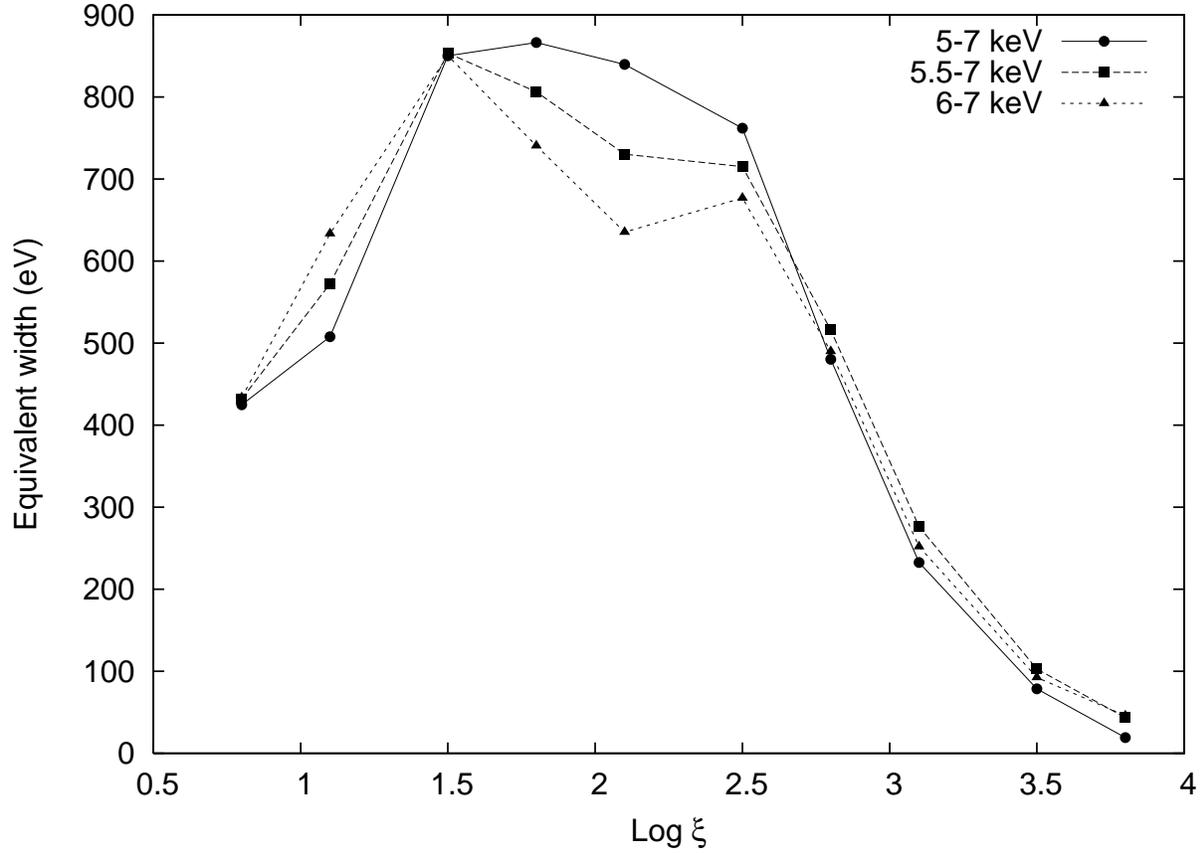}
\caption{Equivalent width of the Fe K$\alpha$ line as a function of the log of the
ionization parameter, resulting from three different energy ranges of integration. 
Connected circles, squares and triangles correspond to $E_{low}=5, 5.5$ and $6$~keV,
respectively. $E_{high}$ is set equal to 7~keV for all the cases.}
\label{f2}
\end{figure*}

Figure~\ref{f2} shows the resulting equivalent widths
as a function of the ionization parameter, calculated for the Fe~K$\alpha$ emission profile
in 10 of our models. Circles connected with solid lines correspond to the integration range
5-7~keV, squares connected with dashed lines to 5.5-7~keV, and triangles connected with
short-dashed lines to the integration performed over the 6-7~keV energy range. It is 
clear from the Figure that the equivalent widths are almost unaffected by the variations
of the lower boundary of the integration range, with the exception of those models with
$1.5 < $log~$\xi < 2.5$. These are the most sensitive cases probably because of the complexity
of the iron K emission. For lower values of the ionization parameter, the emission
mainly occurs at 6.4~keV due to mostly neutral Fe ions. Higher values of $\xi$ means that
the gas is highly ionized and thus mostly He- and H-like Fe ions are responsible for
the emission at $\sim 6.9$~keV. In both cases the line profile is simple in the sense
that the emission is concentrated at one particular energy. In between, emission from
many different ions takes place at the same time, creating a more complex spectral 
feature, as can be seen in the spectrum for log~$\xi=1.8$ in Figure~\ref{f1}.
Since there is no clear reason to choose one of these values of $E_{low}$, we choose
the intermediate one ($E_{low}=5.5$~keV). We consider the uncertainties in the equivalent
widths to be of order of the differences between the values shown in Figure~\ref{f2} 
($\le 100$~eV).

\begin{figure*}
\epsscale{1.0}\plotone{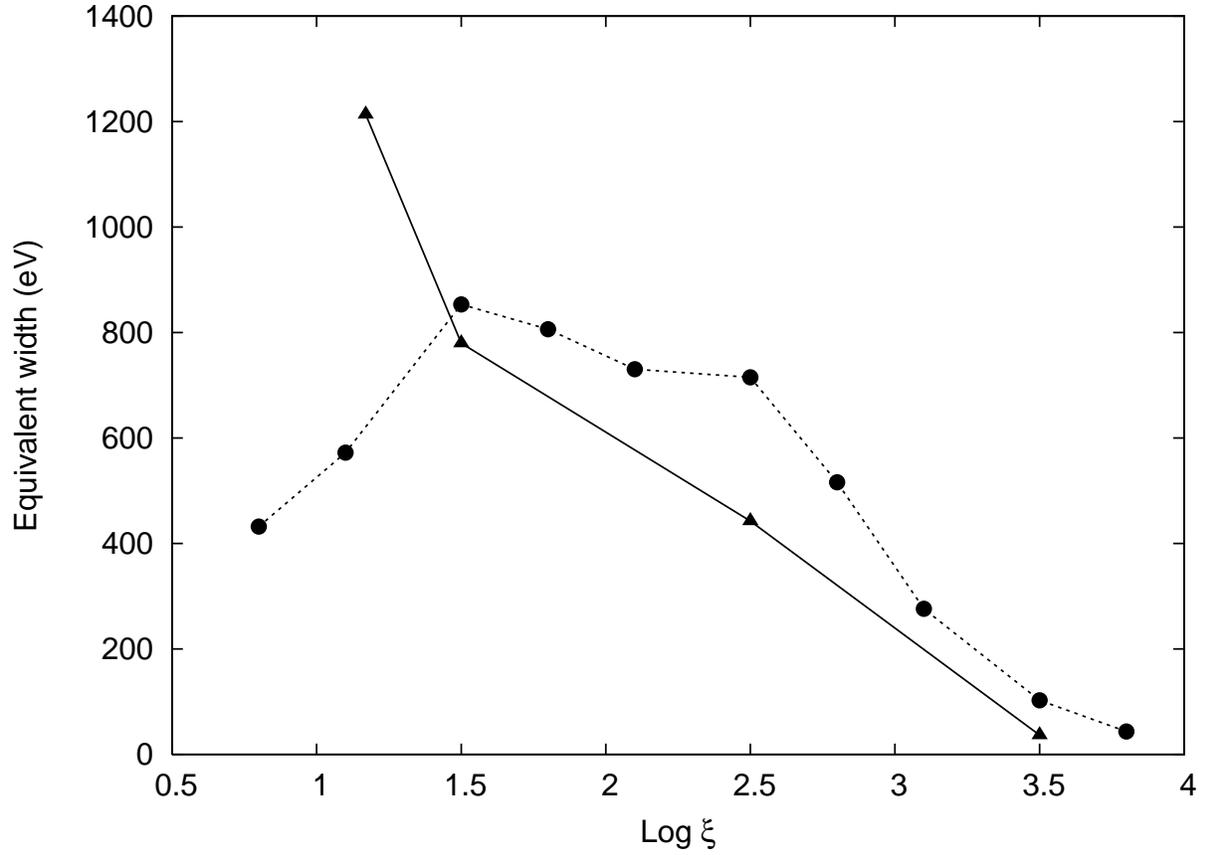}
\caption{Equivalent width of the Fe K$\alpha$ line as a function of the log of the
ionization parameter. Connecting circles are the values predicted by our reflection
code {\sc xillver}, while triangles correspond to the values predicted by {\sc reflion}
\citep{ros05}. The integration is performed over the 5.5-7~keV energy range in both
models.}
\label{f3}
\end{figure*}

Figure~\ref{f3} shows a comparison of the Fe K$\alpha$ equivalent widths predicted
by our models (circles connected with dashed lines), and those predicted by the models 
included in {\sc reflion} \citep[triangles connected with solid lines,][]{ros05}. 
The calculation of the equivalent widths is the same in both models, with the integration
performed in the 5.5-7~keV energy range. There are important differences to notice in
this comparison. It is convenient to make the distinction between two regions in
the plot; the region for models with log~$\xi > 1.5$ and the one for models with log~$\xi < 1.5$.
In the high ionization region, both sets of models show a similar behavior, the equivalent
widths decrease monotonically as the ionization parameter increases, resembling the
Baldwin effect for X-rays. Nevertheless, in this region all our models systematically
predict stronger Fe K$\alpha$ emission. 
For those models in the low ionization region (log~$\xi < 1.5$), the differences are
significantly larger. In fact, {\sc reflion} models predict that the equivalent widths
keep growing as the ionization parameter decreases, while our models show the opposite 
trend. This turn over in the values of the equivalent widths can be understood by looking the ionization 
balance in detail. \cite{kal04} performed similar calculations in photoionized models
using the same atomic data used in our models. Using a set of {\sc xstar} models
of thin spherical shells illuminated by the same power law spectrum they calculated the
ratio of the emissivity per particle for K line production as a function of the
ionization parameter. Their Figure~7 shows the contribution of each ion of iron summing
over the K lines for each one. The overall behavior of these models is quite similar
to the ones presented in this paper. The combined line emissivities have a minimum
just below log~$\xi\sim 1$ (roughly where our calculations start), and is mainly due
to a combination of Fe~{\sc xiv-xvi}. As the ionization increases the also the emissivities
increase, with maximum values that peak around log~$\xi\sim 2$, where the emission
is dominated by Fe~{\sc xviii} and Fe~{\sc xix}. For log~$\xi > 2.5$ the overall line 
emission decreases rapidly as the iron ions become more ionized. This clearly resembles 
the general trend exhibit by the equivalent widths derived from our models, as is shown 
in Figure~\ref{f3}. It is worthwhile to mention that the equivalent widths will increase
again for log~$\xi < 1$, as can be seen in Figure~7 of \cite{kal04}, but those values of
the ionization parameter are out of the range of the calculations presented here. The sensitivity
of the iron K$\alpha$ equivalent width to the ionization parameter represents an important
potential diagnostic of photoionized plasmas, given its ubiquity in the observed 
spectra from accreting sources.

\begin{figure*}
\epsscale{1.0}\plotone{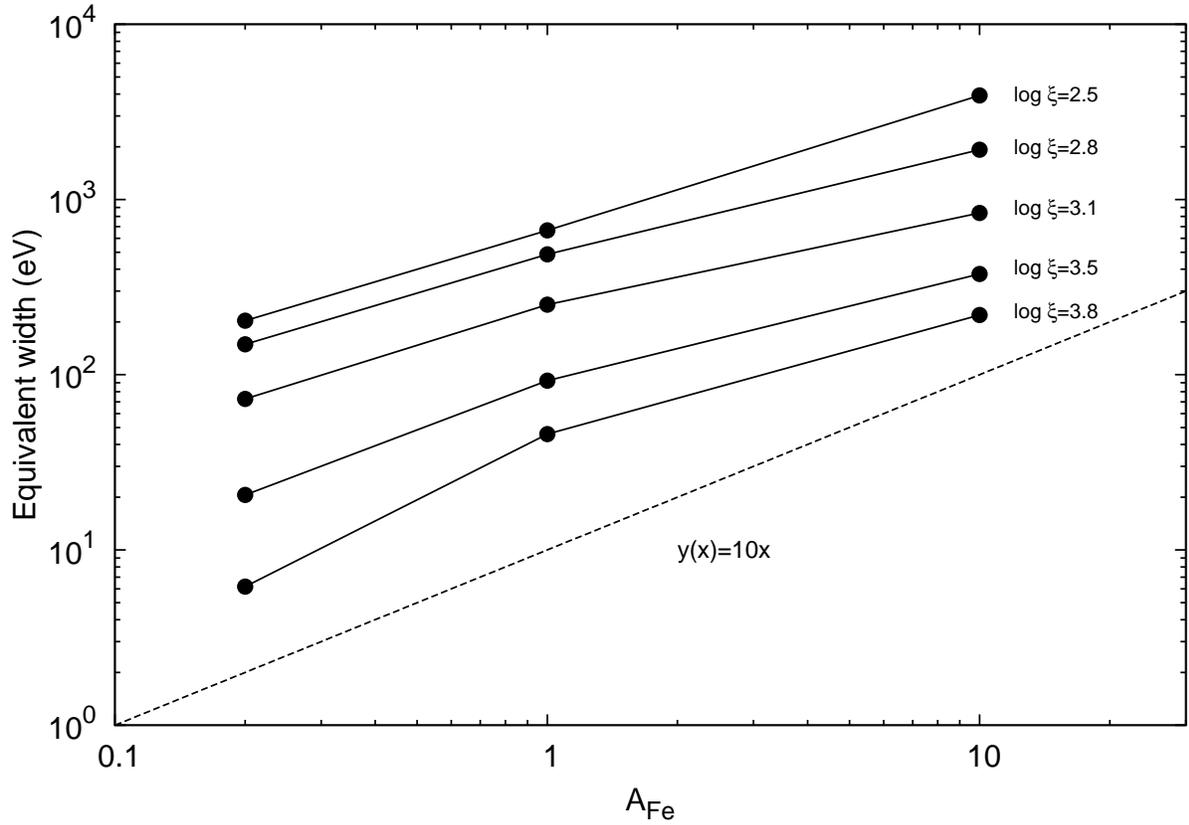}
\caption{Fe K$\alpha$ equivalent width as a function of the iron abundance 
(with respect to solar). Connecting circles are the values predicted by {\sc xillver}
for different illuminations. The value of log~$\xi$ is shown next to each curve. The
dashed line represents the function $y(x)=10x$, to exemplify the linear dependency
of these quantities.}
\label{f4}
\end{figure*}

Another important diagnostic can be achieved through the dependence of the equivalent
width of the K$\alpha$ line of iron on its abundance with respect to the solar value.
This can be seen in Figure~\ref{f4}, where the equivalent width resulting from models
with different iron abundances is plotted. In the Figure, connecting lines correspond 
to a particular ionization parameter (shown next to each curve). For each value of $\xi$,
three reflected spectra were calculated assuming 0.2, 1 and 10 times the solar abundance
of iron. The resulting equivalent widths are shown as filled circles for each case.
It is clear from Figure~\ref{f4} that the dependence of the Fe K$\alpha$ equivalent width
on the ionization parameter is almost the same for different abundances of iron, i.e.,
the shape of the curve in Figure~\ref{f3} will remain unchanged if we use models with
iron over or under abundant, and only the actual values of the equivalent widths will
show a variation. This is a consequence of the fact that the Fe K$\alpha$ equivalent
width grows linearly with the iron abundance, which is also evident from the Figure.
To illustrate this, we include the function $y(x)=10x$ with a dashed line in Figure~\ref{f4},
which shows a slope similar to all the other curves.

\subsection{Simulated observations}\label{secsim}
The analysis presented previously provide important tools for the interpretation
of the X-ray spectra from accreting sources by the use of simple quantities that
can be easily derived from the observed spectrum. Nevertheless, the models computed 
here do not necessarily represent real observed data. First, the spectral
resolution used in the models ($R\sim 350$), although comparable to grating spectra,
is far superior to what is achieved in CCD observations. Second, the models 
do not suffer from loss of information due to the natural restrictions involved in
real observations, such as the instrument effective area which affects the total number
of photons collected. Finally, a real observation is, in general, a mixture of several
components and not exclusively represented by a reflected spectrum. In accreting 
sources such as active galactic nuclei (AGN) and galactic black holes (GBHs), 
an important component is the contribution of the 
source of X-rays that illuminates the accretion disk. In fact, it is not clear 
which fraction of the direct over the reflected component can be present in a given
observation, and this proportion is most likely to depend on the geometry and 
orientation of the system.

\begin{figure*}
\epsscale{1.0}\plotone{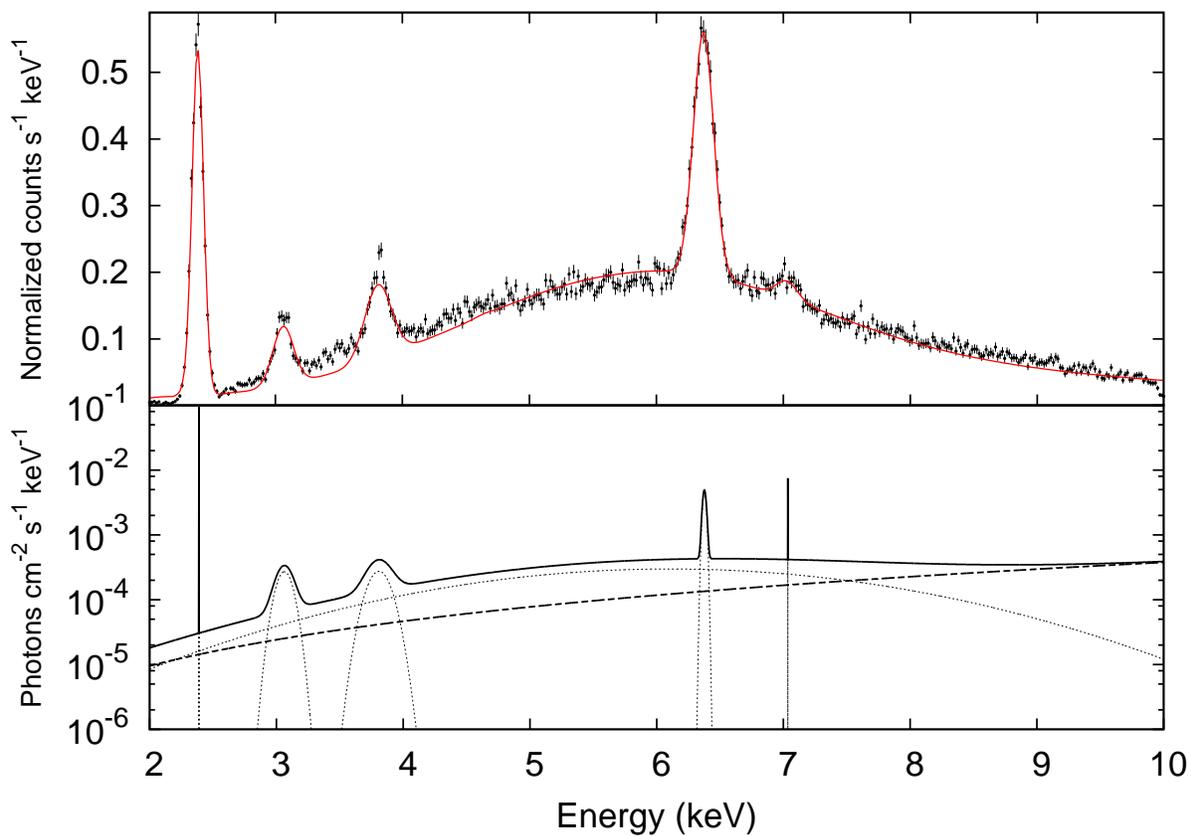}
\caption{Upper panel: data points are the simulated CCD observation produced with our model
for log~$\xi=0.8$. The solid line is the best-fit using a simple phenomenological 
model. Error bars are shown for all the data points. Lower panel: 
the theoretical model used in the fit is shown with the solid 
line, while all the individual components are shown with dashed curves for the 
Gaussian profiles, and with a thick-dashed line for the power-law.}
\label{f5}
\end{figure*}

In order to account for some of these effects, we have produced simulated CCD
observations using our models as the source of the reflected spectra. We have used
the {\it fake it} task in {\sc xspec} in combination with the FI XIS response 
matrices from {\it Suzaku} to simulate 100 ksec observations, assuming a source
flux of $3\times 10^{-10}$ erg cm$^{-2}$ s$^{-1}$ in the 2-10~keV energy band.
This flux corresponds to 10~mCrab, which is a typical value for bright AGN.
We then try to fit the resulting simulated observation with a simple phenomenological
model, i.e., the basic continuum is fitted with a power law, and all the relevant
features with Gaussians. An example of such a fit is shown in Figure~\ref{f5}.
In the upper panel, data points with error bars represent the simulated data using a 
reflection model with log~$\xi=0.8$, while the solid line shows the best-fit.
The lower panel shows the theoretical model used with all its components. The 
continuum is fitted with a power law (thick dashed line), and several emission
profiles with Gaussians (thin dotted lines) centered at 2.39, 3.06, 3.8, 6.38
and 7.03~keV, corresponding to emission from S~{\sc vi-x}, Ar~{\sc viii-xi},
Ca~{\sc x}, Fe~{\sc x-xii} K$\alpha$ and Fe K$\beta$, respectively. Finally,
a very broad ($\sigma=1.54$~keV) Gaussian profile centered at 6.1~keV was included in 
order to properly fit the continuum at lower energies. Clearly, a simple power
law cannot describe the continuum for models with low ionization, given the
significant modification of the original power law continuum due to the large
values of the photoelectric opacity. This very broad curving continuum is required
to fit the reflected spectra with log~$\xi\sim 1.1$ or lower.
Therefore, this particular profile could be used as a signature of strong
reflection from low ionized atmospheres.

We have then produced simulated observations using all the models considered
previously, i.e., using the 10 reflected spectra calculated for log~$\xi=0.8, 1.1,
1.5, 1.8,...,3.8$. By including the {\it Suzaku} response matrices we can 
simulate the effects of the effective area from the instrument, its sensitivity
and energy resolution. However, many X-ray observations from accreting sources
reveal that the direct, unprocessed radiation as a power-law is detected 
simultaneously along with the reflected component. Therefore, we have produced
additional simulated observations with a source spectrum resulting from the 
combination of both the reflected and the illuminating components. In order
to quantify the proportion of one component to the other we use a $\beta$ 
parameter, defined as the ratio of the incident over the reflected flux,
\begin{equation}\label{eqbeta}
\beta = \frac{F_{\mathrm{inc}}}{F_{\mathrm{ref}}} 
\end{equation}
where both fluxes are defined in the 2-10~keV energy range. Note that this
quantity differs from the reflection fraction $R=\Omega/2\pi$, where $\Omega$ 
is the solid angle sustained by the reflector. The relation between the reflection
fraction and our $\beta$ parameter is roughly $R\approx 1/\beta$; such that
a {\it pure reflection} case corresponds to $\beta=0$ and $R\gg 1$, while
in the case of {\it no reflection} $\beta \gg 1$ and $R=0$. Because our models
do not carry any information about the geometry of the reflector, it is not
appropriate to derive a reflection fraction $R$ for them. Instead, we use the
$\beta$ parameter to quantify the dilution of the spectral profiles by the 
direct power-law spectrum, in order to compare derived spectral quantities
with the those from real observations.

\begin{figure*}
\epsscale{1.0}\plotone{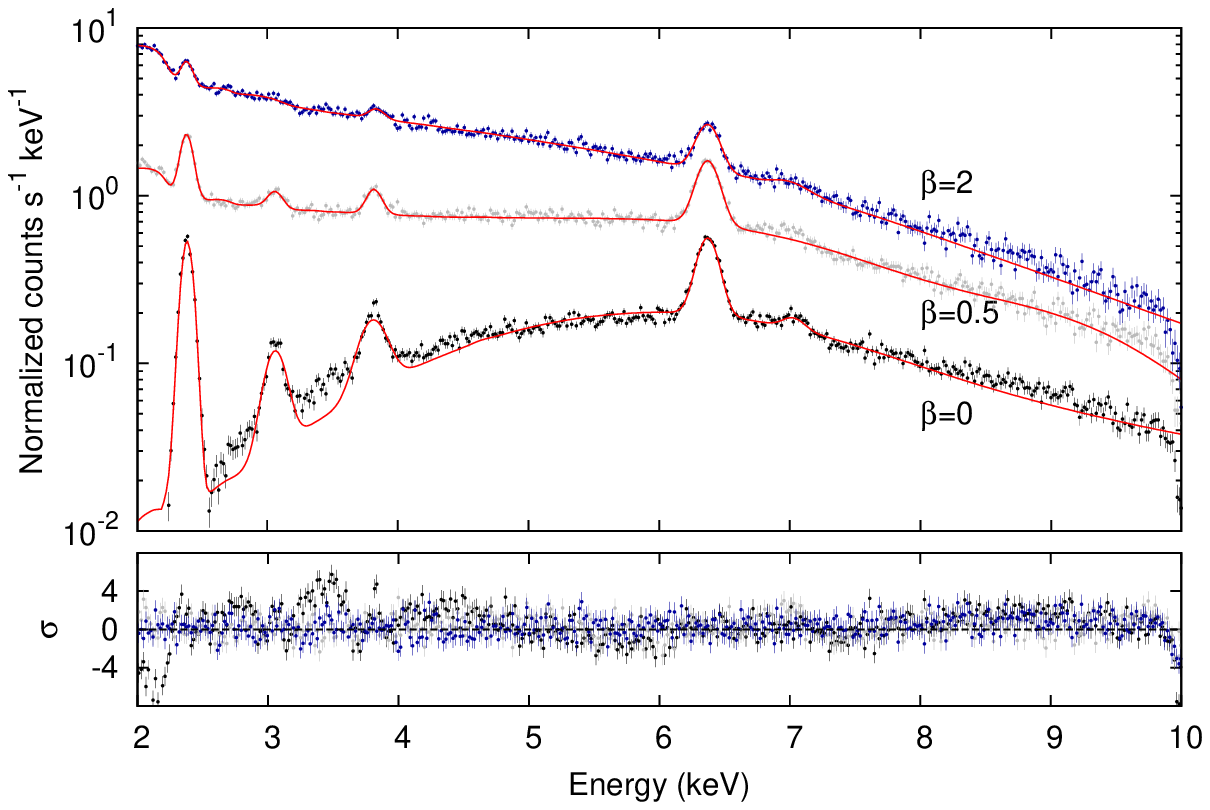}
\caption{Upper panel: data points show the simulated CCD spectra produced with our model
for log~$\xi=0.8$, for $\beta=0, 0.5$ and $2$, from bottom to top. The solid lines
are the best-fits for each case. Lower panel: residuals in units of $\sigma$ for 
all the fits. Error bars are shown for all the data points.}
\label{f6}
\end{figure*}

\begin{figure*}
\epsscale{1.0}\plotone{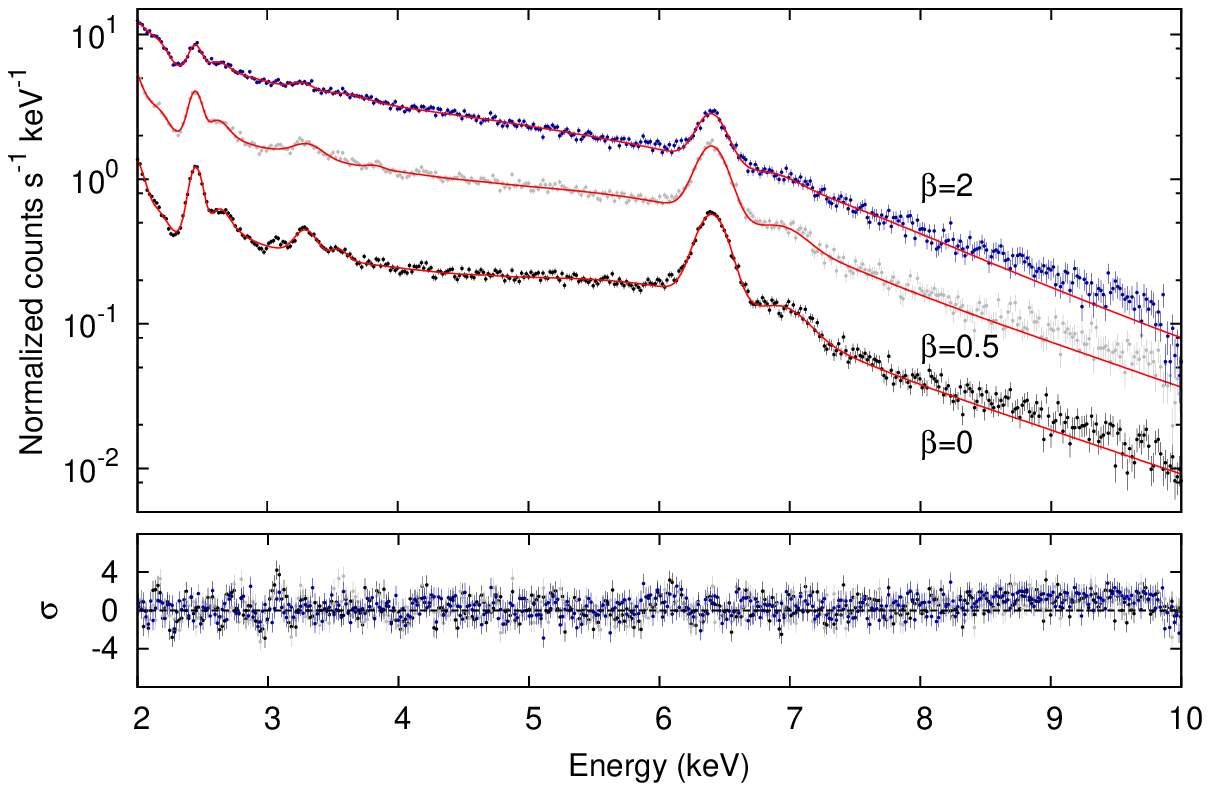}
\caption{Upper panel: data points show the simulated CCD spectra produced with our model
for log~$\xi=1.8$, for $\beta=0, 0.5$ and $2$, from bottom to top. The solid lines
are the best-fits for each case. Lower panel: residuals in units of $\sigma$ for
all the fits. Error bars are shown for all the data points.}
\label{f7}
\end{figure*}

\begin{figure*}
\epsscale{1.0}\plotone{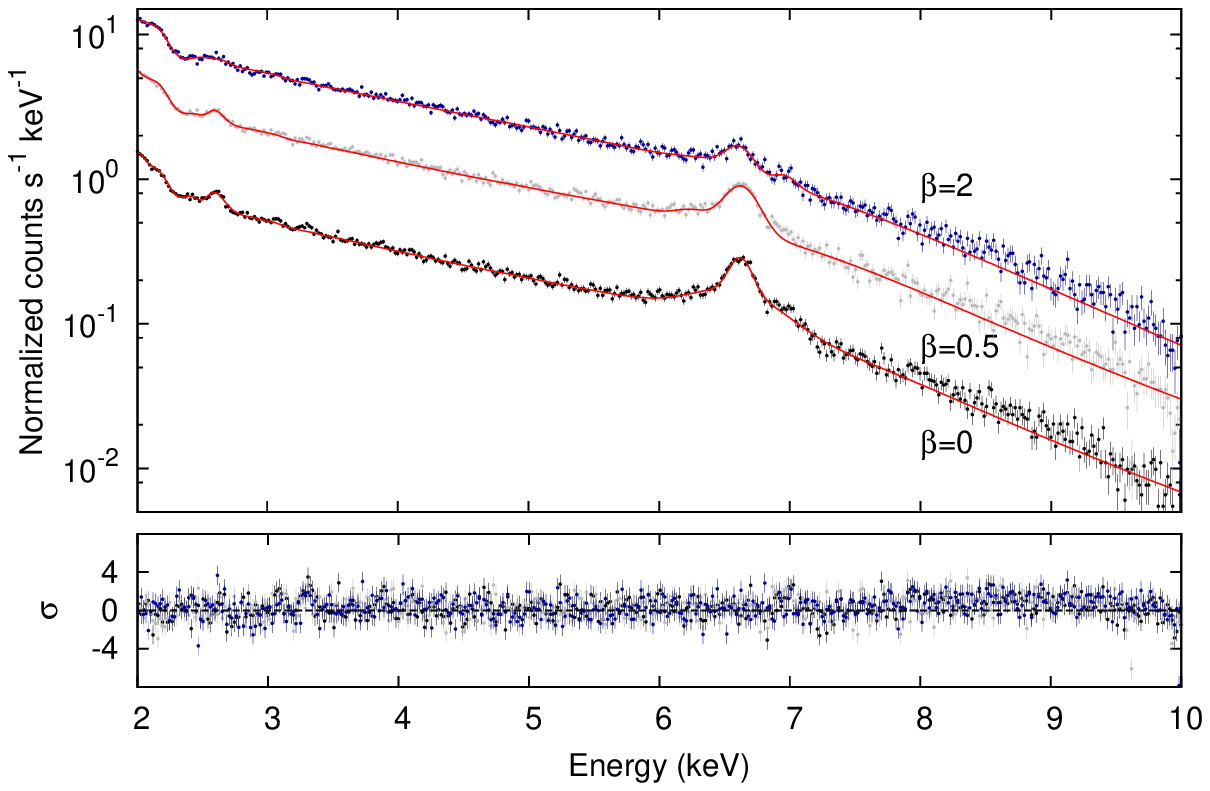}
\caption{Upper panel: data points show the simulated CCD spectra produced with our model
for log~$\xi=2.8$, for $\beta=0, 0.5$ and $2$, from bottom to top. The solid lines
are the best-fits for each case. Lower panel: residuals in units of $\sigma$ for
all the fits. Error bars are shown for all the data points.}
\label{f8}
\end{figure*}

Figures~\ref{f6}, \ref{f7} and \ref{f8} show the simulated CCD observations
using our models with log~$\xi=0.8, 1.8$ and $2.8$, respectively. In each figure,
the data points in the upper panel show the simulated data for three different values 
of the $\beta$ parameter, namely, from bottom to top, $\beta=0, 0.5$ and $2$.
The solid lines are the best-fit for each case, using the same kind of simple
models described before. The dots in the lower panel represent the residuals
in units of $\sigma$ for all the fits. These figures show the effect of the 
instrument on the observed spectra. In particular, by comparing the curve for
$\beta=0$ in Figure~\ref{f6} with the corresponding model in Figure~\ref{f1},
it is clear how the narrow Fe K$\alpha$ and K$\beta$ lines in the log~$\xi=0.8$ 
model become much broader features in the simulated observation. As mentioned
before, emission from S, Ar and Ca ions are the most prominent features
at energies $< 6$~keV. It is also important to notice from this Figure the 
dilution of the spectral profiles due to the inclusion of the direct power-law
spectra (i.e., the curves for $\beta=0.5$ and $\beta=2$). The iron 
K$\alpha$ at 6.4~keV and the sulfur blend at $\sim 2.39$~keV emission profiles
are clearly detectable, even for $\beta=2$ (i.e., twice as much direct flux
as the reflected flux in the 2-10~keV band). However, the weaker emission 
lines become almost undetectable for $\beta \ge 2$. Most important, not only
the atomic features are modified but the continuum as well. The broad Gaussian
profile required to fit the continuum in combination with the power-law
is no longer required for $\beta=0.5$ or greater. This is important, since
this means that this feature is characteristic of low ionization, but also
appears exclusively for the pure reflection cases ($\beta=0$).

Figures~\ref{f7} and \ref{f8} show essentially the same situation. In 
Figure~\ref{f7}, which corresponds to the simulation generated with 
log$\xi=1.8$, it is important to notice the presence of the S~{\sc xv} K$\alpha$ 
emission line at $\sim 2.45$~keV, and the S~{\sc xvi} K$\alpha$ plus the 
radiative recombination continua (RRC) at 
$\sim 2.59$~keV, which are detectable even in the $\beta=2$ simulated spectrum.
The iron K$\beta$ emission line becomes undetectable for $\beta > 1$.
Figure~\ref{f8} shows the log$\xi=2.8$ simulated spectra. In this case, the
iron emission is the only clear and strong feature. However, the iron K-shell
emission can no longer be well represented by a single Gaussian profile. Instead,
a broad ($\sigma=1.26$~keV) Gaussian centered at $\sim 6.5$~keV accounts for the Comptonized
profile, while a narrow one centered at $\sim 6.6$~keV accounts for the
discrete emission lines that can be seen in the original spectrum of Figure~\ref{f1}.
The iron emission can be well fitted even in the $\beta=2$ case, but not for
cases with a larger contribution of the direct power-law spectrum.

In Table~\ref{tafea} we show a compilation of the strongest features detectable
in the simulated CCD spectra for the pure reflection case ($\beta=0$). We show
the line centroid energy and the equivalent with for each feature. Three of the
features are blends of several components, such as the Si~{\sc vi-x} around $2.39$~keV,
the blend of the S~{\sc xi} K$\alpha$ line and the Si~{\sc xiv} RRC around $2.59$~keV, i
and the Ar~{\sc viii-xi} blend around $3.3$~keV. 

\subsection{Comparison with observations}\label{seccomp}
In order to test the validity of the results obtained with the simulated CCD spectra,
we have compared our estimates with those derived from real observations. To do this
we have concentrated our attention on the equivalent widths and centroid line energies
of the iron K$\alpha$ emission profile, since this is the strongest emission present in 
the spectra from most accreting sources. Because the resulting equivalent width is very
sensitive to small changes in the fitted continuum, we decided to apply a systematic 
procedure to calculate the line equivalent width, similar to the one described in 
Section~\ref{secews}. First, we have generated simulated CCD spectra for the 10 different
values of the ionization parameter considered in this paper (log~$\xi=0.8, 1.1, 1.5, 1.8,...,3.8$),
and for 5 different values of the $\beta$ parameter from Equation~\ref{eqbeta} 
($\beta=0, 1, 2, 5, 10$), which covers cases from {\it pure reflection} cases up to
cases dominated by the direct power-law ({\it no reflection}). Then, we fit a simple model
to each spectrum, consisting of a power-law for the local continuum and one Gaussian
profile for the Fe K-shell emission. The fit is performed in a small energy range
concentrated around the iron lines, ignoring any other channels. We choose the 
range to be 5.7-7.3~keV for spectra with $0.8 \ge \mathrm{log}~\xi \ge 2.8$, and
3-10~keV for those with larger ionization parameters, since the Fe K-line is
much broader due the Comptonization.

\begin{figure*}
\epsscale{1.0}\plotone{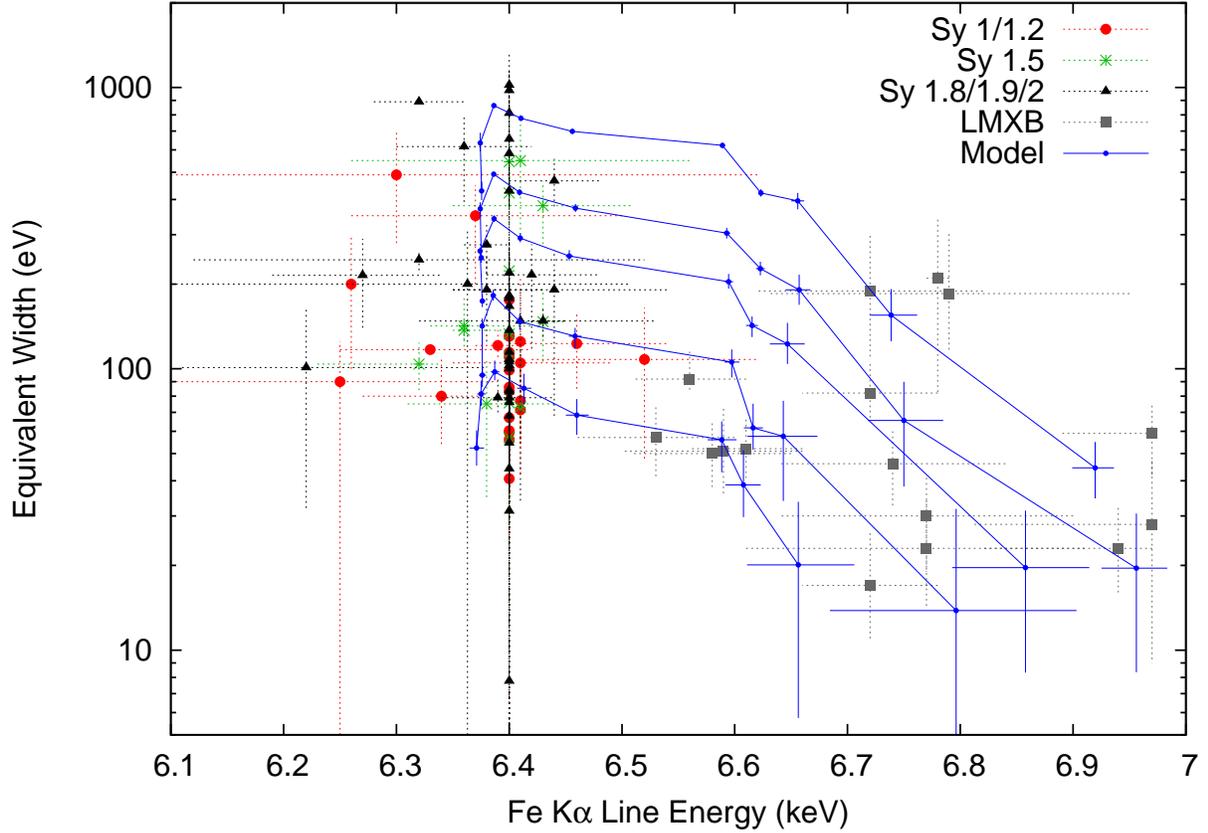}
\caption{Equivalent width versus centroid energy of the Fe K$\alpha$ line. Filled circles,
asterisks and filled triangles correspond to AGN Seyfert 1-1.2, 1.5 and 1.8-2 galaxies,
respectively \citep{win09}. Filled squares correspond to LMXB \citep{ng10}. Dots connected 
with solid lines are the values predicted by our models with $0.8\le \mathrm{log}~\xi \le 3.8$
(from left to right). From top to bottom, each curve corresponds to $\beta=0, 1, 2, 5$ and $10$. 
Error bars are shown for both data and model points.}
\label{f9}
\end{figure*}

In Figure~\ref{f9}, the line centroid energies and equivalent widths resulting from 
all our simulated spectra are compared with the same quantities derived from spectra 
of Seyfert (Sy) galaxies and low mass X-ray binaries (LMXB). In the Figure, filled
circles correspond to Sy 1 and 1.2 galaxies, asterisks to Sy 1.5, and filled triangles
to Sy 1.8, 1.9 and 2; all taken from the \cite{win09} compilation. Filled squares
correspond to the LMXB analyzed by \cite{ng10}. The error bars are included as dashed
lines. Dots connected with solid lines are the values predicted by our reflection
models. Each line corresponds to one particular ratio of the direct over the reflected
component, or $\beta$ parameter. From top to bottom, each curve corresponds to 
$\beta=0, 1, 2, 5$ and $10$. Error bars are also included as solid lines. Along each 
line, the points from left to right correspond to the models with increasing values
of the ionization parameter $\xi$.

There are important pieces of information contained in Figure~\ref{f9}. The observed
data shows the clear distinction between Sy galaxies and LMXB with respect to their
iron emission. Sy galaxies seem to only show the {\it cold} iron line at $\sim 6.4$~keV,
while in the LMXB spectra the {\it hot} line around 6.7~keV is detected. This yields
a marked difference in the ionization of the emitting gas between these two types
of sources. According to our models, the cold region of the plot where Sy galaxies
appear (for line energies lower than 6.5~keV), correspond to values of log~$\xi < 2$, 
while the hot region (line energies larger than 6.5~keV), where LMXB show up, covers 
larger values of ionization, up to log~$\xi=4$. In the cold region, Sy galaxies 
equivalent widths cover a large range, from 10 to 1000~eV. Despite the large error
bars in many of the values, equivalent widths larger than 200~eV seem to correspond
almost exclusively to Sy 1.5-2 galaxies. Our models predict that the largest values
correspond to reflection dominated cases, with $\beta=0, 1$ and 2. Lower equivalent
widths (less than 200~keV), seem to be common to all types of Sy galaxies. However,
our models require significant contribution from the direct power-law illuminating
spectra to dilute the line profiles and reproduce the lower values of the equivalent
widths. Accordingly, we need to produce models with $\beta=10$ to explain the equivalent
widths lower than 100~eV (i.e., 10 times more direct power-law flux than the reflected
one, or $R\rightarrow 0$). Also, it is important to notice that for the lower ionization models the resulting
line energies are much closer to each other than for the high ionization ones. At the same time,
the data for Seyfert galaxies in Figure~\ref{f9} shows a large concentration of points
at 6.4~keV. This is probably due to the fact that sometimes observers fix
the line centroid energy to a known value given the poor statistics of the data.
This imposes difficulties in the estimation of the ionization parameter when comparing
the data values to the ones predicted by the models. Therefore, accurate line energies
are needed from the observations in order to improve the diagnostics.

The LMXB observations are more scattered in the hot region of the Figure~\ref{f9}
(line energies larger than 6.5~keV). Both the models the the observed data predict
equivalent widths as small as $\sim 10$~eV, but no larger than $\sim 200$~eV.
The error bars in the values resulting from the models are much larger than in the
low ionization cases, mainly because the line feature is weaker as the ionization
increases, since the emission is mostly due to H- and He-like iron. The line profile
is even weaker as the $\beta$ parameter increases, due to the direct power-law
dilution. Therefore, in the cases for $\beta=1$ and 2 (third and fourth curves from
top to bottom), the line cannot be detected in the spectra simulated with highest
ionization parameter (log~$\xi=3.8$). The same occurs for the simulated spectra
with log~$\xi=3.5$ in the bottom curve ($\beta=10$). The ranges of the ionization
and the $\beta$ parameters covered in this paper are sufficient to well represent 
the observed data for both Seyfert galaxies and LMXB.

\begin{figure*}
\epsscale{1.0}\plotone{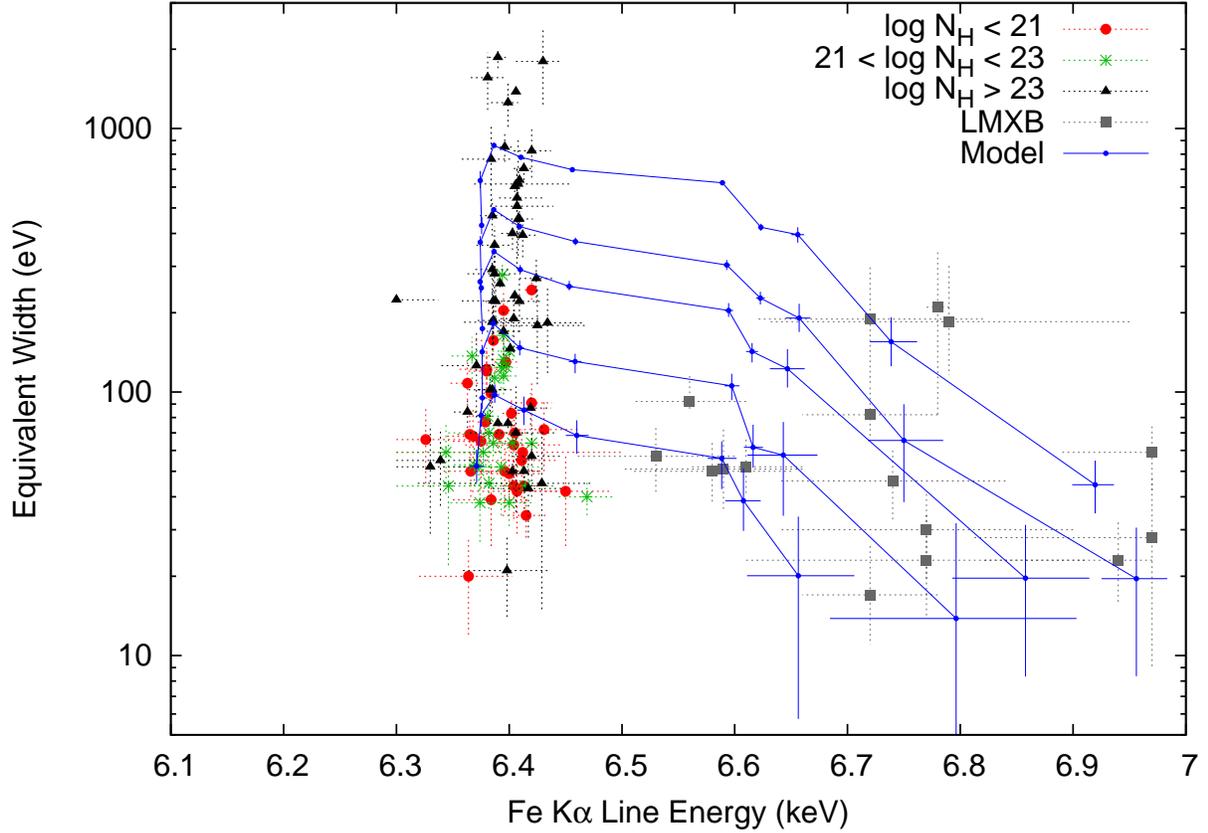}
\caption{Equivalent width versus centroid energy of the Fe K$\alpha$ line. Filled circles,
asterisks and filled triangles correspond to AGN Seyfert galaxies with log~$N_{\mathrm H}<21$,
$21 <$ log~$N_{\mathrm H}<23$, and log~$N_{\mathrm H}>23$~cm$^{-2}$, respectively \citep{fuk10}. 
Filled squares correspond to LMXB \citep{ng10}. Dots connected
with solid lines are the values predicted by our models with $0.8\le \mathrm{log}~\xi \le 3.8$
(from left to right). From top to bottom, each curve corresponds to $\beta=0, 1, 2, 5$ and $10$.
Error bars are shown for both data and model points.}
\label{f10}
\end{figure*}

The fact that the largest equivalent widths predicted by our reflection dominated  
models correspond mostly to the values found in Sy 1.5-2 galaxies agrees with the 
general picture that in these sources the direct component of X-ray radiation is 
being obscured, and only the reflected or reprocessed radiation is observed.
Following this idea, we repeated the comparison of our models with a larger and more
recent compilation of AGN Seyfert galaxies done by \cite{fuk10}. These authors 
analyzed high-quality {\it Suzaku} data for 88 Seyfert galaxies, and derived line
energies, equivalent widths, and hydrogen column densities among other quantities.
In Figure~\ref{f10} we show the equivalent width against the Fe K$\alpha$ line energy 
derived from these observations. For comparison we use the same scale used in 
Figure~\ref{f9}, but in this case we use different
symbols to show different values of the hydrogen column density $N_{\mathrm H}$. 
In the plot, filled circles represent galaxies with log~$N_{\mathrm H} < 21$~cm$^{-2}$,
asterisks are galaxies with $21 \le N_{\mathrm H} < 23$~cm$^{-2}$, and filled triangles are
those with log~$N_{\mathrm H} \ge 23$~cm$^{-2}$. As in Figure~\ref{f9}, filled squares are 
the values for the LMXB from \cite{ng10}, and dots connected with solid lines are
the values predicted by our models. This particular sample has the advantage of being 
more accurate (lower error bars), and is more consistent since all the data was
obtained with the same instrument and analysis technique. However, the information derived from this comparison
is consistent with the \cite{win09} compilation. Equivalent widths larger than 
200~eV are only observed in sources with large hydrogen column densities 
(log~$N_{\mathrm H} \ge 23$~cm$^{-2}$), which corresponds to the reflection-dominated 
models ($0 \le \beta \le 2$). Lower equivalent widths are observed in sources
with all values of $N_{\mathrm H}$. The majority of sources with low hydrogen
column densities (filled circles) have equivalent widths lower than 100~eV,
consistent with models with large contribution of the direct power-law 
($\beta\sim 10$). The agreement between these observations and our models is
in general very good. However, there are at least five galaxies with equivalent
widths larger than the maximum values predicted by our models ($\sim 1$~keV)
with solar abundances. However, larger equivalent widths can be easily achieved
by increasing the iron abundance, as shown in Figure~\ref{f4}.

%
%==================================================================================
%
\section{Conclusions}\label{seccon}
In this paper we have presented a comprehensive study of the emission spectra from
accreting sources. Using our new reflection modeling code \citep{gar10}, we have 
computed the reflected spectra from an X-ray illuminated accretion disk. We have
concentrated our analysis to the iron K-shell emission lines, although other spectral
features are also discussed. Our models predict equivalent widths for the Fe K$\alpha$ 
emission of $\sim 10$~eV for high ionization parameters (log~$\xi\sim 4$), and maximum 
values of $\sim 800$~eV for models with log~$\xi\sim 1.5$. For lower values of the 
ionization parameter the equivalent widths decrease to a minimum near log~$\xi\sim 1$,
contrary to what other models predict. These differences are due to the
atomic data used by each simulation. We have shown that the behavior of the Fe K$\alpha$
equivalent widths with respect to the ionization parameter of the gas is consistent
with the line emissivities shown by \cite{kal04}, where the same atomic data was used.
Additionally, these equivalent widths display a linear dependency with the iron abundance
normalized to its solar values. This seems to be true for all the models within the
range of ionization parameter values considered in this paper.

Simple analysis of simulated CCD spectra reveals that for low ionized, reflection
dominated cases the 2-10~keV continuum cannot be represented by a simple power-law. 
Instead, a broad Gaussian profile is also required. This type of continuum can be 
used as a strong reflection signature. These simulations also indicate that in addition
to the iron K-shell lines, the S~{\sc xv} K$\alpha$ emission line at 2.45~keV is one
of the most prominent features in the spectra. For cases with log~$\xi\sim 2$, this
line blends with the He-like silicon RRC providing a broad feature that cannot be
represented by simple Gaussian profiles. This could also be used as another reflection
signature while analyzing observations.

The iron K$\alpha$ equivalent widths and line centroid energies are in good agreement
with values reported in the literature for both AGN Seyfert galaxies, and galactic sources
such as LMXB. In particular, both model and observations show that large equivalent
widths ($> 200$~eV) can be achieved in many situations. According to our models, the 
largest values of the equivalent widths correspond to the lower $\beta$ parameters, i.e.,
reflection dominated cases. Only observational data from AGN Sy 1.8-2 galaxies, or those 
with log~$N_{\mathrm H} \ge 23$~cm$^{-2}$ coincide with these large equivalent widths, but only 
for low values of the ionization parameter (log~$\xi < 2$). The majority of the Sy 1-1.2,
or those sources with log~$N_{\mathrm H} < 21$~cm$^{-2}$ show equivalent widths below 200~eV.
In order to reproduce those values with our models, we need to increase the contribution 
of the direct power-law up to ten times the contribution of the reflected component.
These results agree with the general idea that type 1 and 2 Seyfert galaxies only
differ on their orientation with respect to the observer. 
In the case of LMXB, both the model and the observed data coincide in equivalent widths 
of $\sim 10-200$~eV. Since all these sources show hotter Fe K-lines (with energies
around $6.7-6.9$~keV), they correspond to models with values of the ionization parameter
larger than those for AGN ($2 < \mathrm{log}~\xi < 4$).

%
%==================================================================================
\acknowledgments
This work was supported by a grant from the NASA astrophysics theory program
05-ATP05-18. This research has made use of NASA's Astrophysics Data System.
%
%==================================================================================
%
\begin{deluxetable}{lrcccccccccc}
\tabletypesize{\scriptsize}
\tablecaption{Strongest features in the simulated CCD spectra
\label{tafea}}
\tablewidth{0pt}
\tablehead{
\colhead{log~$\xi$} & (keV) & \colhead{0.8} & \colhead{1.1} & \colhead{1.5} &  \colhead{1.8} & 
\colhead{2.1} & \colhead{2.5} & \colhead{2.8} & \colhead{3.1} & \colhead{3.5} & \colhead{3.8} \\
}
\startdata
Si~{\sc xiv} K$\alpha$ & $E_{line}$  &      &      &      & 1.98 & 1.93 & 2.01 & 2.01 &      &      &      \\
                       &  EW         &      &      &      & 0.19 & 0.26 & 0.04 & 0.02 &      &      &      \\
Si~{\sc vi-x}          & $E_{line}$  & 2.39 & 2.39 &      &      &      &      &      &      &      &      \\
                       &  EW         & 6.12 & 1.03 &      &      &      &      &      &      &      &      \\
S~{\sc xv} K$\alpha$   & $E_{line}$  &      &      & 2.46 & 2.45 & 2.45 & 2.44 &      &      &      &      \\
+ Si~{\sc xiii} RRC    &  EW         &      &      & 0.36 & 0.29 & 0.17 & 0.03 &      &      &      &      \\
S~{\sc xi} K$\alpha$   & $E_{line}$  &      &      & 2.58 & 2.59 & 2.59 & 2.59 & 2.57 &      &      &      \\
+ Si~{\sc xiv} RRC     &  EW         &      &      & 0.05 & 0.09 & 0.10 & 0.04 & 0.03 &      &      &      \\
Ar~{\sc viii-xi}       & $E_{line}$  & 3.06 & 3.07 & 3.29 & 3.29 & 3.34 &      &      &      &      &      \\
                       &  EW         & 0.63 & 0.10 & 0.07 & 0.11 & 0.23 &      &      &      &      &      \\
Ca~{\sc x}             & $E_{line}$  & 3.81 & 3.83 & 3.84 &      &      &      &      &      &      &      \\
                       &  EW         & 0.43 & 0.14 & 0.03 &      &      &      &      &      &      &      \\
Fe K$\alpha$           & $E_{line}$  & 6.38 & 6.38 & 6.39 & 6.41 & 6.46 & 6.61 & 6.61 & 6.65 & 6.74 & 6.92 \\
                       &  EW         & 0.38 & 0.84 & 1.07 & 0.87 & 0.65 & 0.42 & 0.37 & 0.39 & 0.16 & 0.04 \\
Fe K$\beta$            & $E_{line}$  & 7.03 & 7.01 & 6.96 & 6.97 & 6.97 &      &      &      &      &      \\
\enddata
\end{deluxetable}
%
%==================================================================================
%
%==================================================================================
%
%==================================================================================
%
%==================================================================================
%
%
%==================================================================================
%
%
%==================================================================================
%
%
%==================================================================================
%
%
%==================================================================================
%
%
%==================================================================================
%
%
%==================================================================================
%
%
%==================================================================================
%
\bibliographystyle{apj}
\bibliography{my-references}
%
%==================================================================================
\end{document}